\documentclass[conference,compsoc]{IEEEtran}
\usepackage{caption}
\usepackage[table]{xcolor}
\usepackage{tikz}
\usepackage{amsmath}
\usepackage{makecell}
\usepackage{xspace}
\usepackage{booktabs} 
\usepackage{graphicx}
\usepackage{tabularx}
\usepackage{float}
\usepackage{algpseudocode}
\usepackage{subfigure}
\usepackage{makecell} 
\usepackage{multirow}
\usepackage{cite}
\usepackage{enumitem}

\graphicspath{ {./images/} }
\usepackage{numprint}
\usepackage{etoolbox}
\usepackage{pgfgantt}
\usepackage{rotating}
\usepackage{tabularray}
\usepackage{hyperref}
\usepackage{cleveref}
\usepackage{comment}
\usepackage{algorithm}
\usepackage{algpseudocode}
\usepackage{amsmath}
\usepackage{adjustbox}

\usepackage{listings}
\usepackage{array}
\definecolor{lightgray}{gray}{0.9}

\definecolor{darkgreen}{rgb}{0.0, 0.5, 0.0}

\newcommand{\etal}{\emph{et~al.}\xspace}
\newcommand{\ournameNoSpace}{\mbox{CAL}}
\newcommand{\ourname}{\ournameNoSpace\xspace}

\newcommand{\sys}{\ourname}
\newcommand{\sysNoSpace}{\mbox{CAL}}

\newcommand{\sect}{Sect.~}
\newcommand{\fig}{Fig.~}
\newcommand{\tab}{Tab.~}
\newcommand{\hsect}[1]{\hyperref[#1]{\sect\ref{#1}}}
\newcommand{\hfig}[1]{\hyperref[#1]{\fig\ref{#1}}}
\newcommand{\htab}[1]{\hyperref[#1]{\tab\ref{#1}}}

\usepackage{tikz}

\newcommand{\CHone}{CH1}
\newcommand{\CHtwo}{CH2}
\newcommand{\CHthree}{CH3}
\newcommand{\CHfour}{CH4}

\newcommand{\CHoneRef}{\hyperref[ch:challenge1]{\CHone}}
\newcommand{\CHtwoRef}{\hyperref[ch:challenge2]{\CHtwo}}
\newcommand{\CHthreeRef}{\hyperref[ch:challenge3]{\CHthree}}
\newcommand{\CHfourRef}{\hyperref[ch:challenge4]{\CHfour}}

\newcolumntype{H}{>{\setbox0=\hbox\bgroup}c<{\egroup}@{}}

\begin{document}

\date{}

\title{GNN-Based Code Annotation Logic for Establishing Security Boundaries in C Code}

\author{
    \IEEEauthorblockN{Varun Gadey\IEEEauthorrefmark{1}, Raphael Götz\IEEEauthorrefmark{1}, Christoph Sendner\IEEEauthorrefmark{1}, Sampo Sovio\IEEEauthorrefmark{2}, Alexandra Dmitrienko\IEEEauthorrefmark{1}}
    \IEEEauthorblockA{\IEEEauthorrefmark{1}University of Würzburg, Germany\\
    }
    \IEEEauthorblockA{\IEEEauthorrefmark{2}Huawei Technologies, Finland\\
    }
}

\maketitle

\begin{abstract}
Securing sensitive operations in today's interconnected software landscape is crucial yet challenging. Modern platforms rely on Trusted Execution Environments (TEEs), such as Intel SGX and ARM TrustZone, to isolate security-sensitive code from the main system, reducing the Trusted Computing Base (TCB) and providing stronger assurances. However, identifying which code should reside in TEEs is complex and requires specialized expertise, which is not supported by current automated tools. Existing solutions often migrate entire applications to TEEs, leading to suboptimal use and an increased TCB. 

To address this gap, we propose Code Annotation Logic (CAL), a pioneering tool that automatically identifies security-sensitive components for TEE isolation. CAL analyzes codebases, leveraging a graph-based approach with novel feature construction and employing a custom graph neural network model to accurately determine which parts of the code should be isolated. CAL effectively optimizes TCB, reducing the burden of manual analysis and enhancing overall security. 
Our contributions include the definition of security-sensitive code, the construction and labeling of a comprehensive dataset of source files, a feature-rich graph-based data preparation pipeline, and the CAL model for TEE integration. Evaluation results demonstrate CAL's efficacy in identifying sensitive code with a recall of 86.05\%, an F1 score of 81.56\%, and an identification rate of 91.59\% for security-sensitive functions. 

By enabling efficient code isolation, CAL advances the secure development of applications using TEEs, offering a practical solution for developers to reduce attack vectors.

\end{abstract}

\section{Introduction}
\label{sec:introduction}

In today’s interconnected software landscape, protecting sensitive operations within applications has become a critical challenge. Although modern computing platforms come equipped with advanced security features like isolation, access control, and memory protection, these measures fundamentally depend on the integrity of the underlying software layer—typically an operating system or hypervisor. This underlying software forms the core of the Trusted Computing Base (TCB), the set of components that must remain secure for the system to function safely. Consequently, any compromise of this software layer endangers the security of all applications and data within the system.

However, operating systems and hypervisors are vulnerable to a variety of attack vectors, including rootkits~\cite{kim2012rootkit}, kernel exploits~\cite{CVE-2024-39291, CVE-2016-5195}, hypervisor vulnerabilities~\cite{CVE-2023-34322,CVE-2023-4155}, and malicious alterations during installation~\cite{CVE-2024-3094,CWE-506}. Such compromises allow attackers to bypass platform security measures altogether, enabling them to manipulate memory~\cite{CWE-119}, extract sensitive data~\cite{CWE-200}, alter control flows~\cite{CWE-691}, disable security checks~\cite{CWE-693}, and inject malicious code into applications at runtime~\cite{CWE-94}. This issue is exacerbated by the inherent complexity and size of modern system software, which make it nearly impossible to formally verify or exhaustively test against all potential vulnerabilities.

A promising approach has emerged to address the growing risk of system software compromise: Dividing applications into security-sensitive and non-sensitive components and executing only the sensitive components under the protection of Trusted Execution Environments (TEEs). TEEs are enabled by hardware-security extensions that provide hardware-based isolation for encapsulating sensitive code and data. Two prominent examples of TEEs are Intel SGX~\cite{intelSGX} and ARM TrustZone~\cite{armTrustZone}, each employing distinct models suited to their hardware architectures.

Intel SGX enables the creation of isolated enclaves, designated sections of an application that execute sensitive code independently of the main system. This protects enclave-resident code and data even from the operating system and hypervisor, ensuring that these components remain secure against privileged software attacks. Although SGX can technically support entire applications — or even operating systems — within an enclave, this increases the TCB, raising the risk of exploitable vulnerabilities. From a security perspective, keeping the TCB within an enclave as small as possible is generally preferable, including only the most critical operations.

ARM TrustZone divides the system into two worlds: Secure and normal. Security-critical operations run in the secure world, isolated from non-sensitive processes in the normal world. Switching between worlds is managed via the Secure Monitor Call (SMC), which invokes the Secure Monitor~\cite{pinto2019demystifying}. The Secure Monitor ensures secure state transitions, protecting memory and registers from data leakage. Resources in the secure world are limited, with memory varying from a few to tens of megabytes, depending on device configuration and hardware capabilities~\cite{armTrustZoneCortexA, armTrustZone}.

For applications that leverage TEEs, the TCB is confined to the code within the enclave (as in Intel SGX) or to the Secure Monitor and code operating in the isolated secure world (as in ARM TrustZone). By narrowing the TCB to these isolated components, TEEs provide stronger security assurances for critical operations, reducing the risk of vulnerabilities in system software and creating a more robust foundation for protecting sensitive data.

However, integrating a TEE into an existing application is complex and requires specialized security expertise. Determining which components are genuinely security-sensitive and which can remain in the regular execution environment involves a labor-intensive analysis of the entire code base. Even the definition of "security-sensitive code" can be challenging, as it often depends on the specific context of the application and the threats it faces. As a result, automating this identification process remains an open challenge, and no fully automated tools currently exist to assist developers in this task. This highlights the critical need for developer teams to have a deep understanding of their applications and the associated security risks. The situation underscores the urgent demand for reliable identification tools that empower developers to utilize TEEs effectively while maintaining robust security. Such tools could alleviate the burden of manual analysis, minimize the risk of oversights, and ultimately enhance the overall security posture of applications leveraging TEEs.

Given the lack of automated tools for code-splitting, many developers move entire applications into the secure environment, aided by tools like Graphene~\cite{tsai2017graphene}, Haven~\cite{baumann201haven}, and SCONE~\cite{arnautov2016scone}. Graphene and Haven enable legacy applications to run in secure enclaves, with Haven focusing on cloud environments. SCONE secures containerized applications, specializing in microservices. These tools automate the migration of \textit{full} applications into TEEs, which leads to introducing unnecessary code into TCB, leading to the blown up TCB and the increased risk of vulnerabilities. 
Only a few approaches~\cite{liu2020reducing, gudka2015clean} have explored the possibility of automatically splitting the code in the context of TEE integration~\cite{liu2020reducing}, or when employing code compartmentalization to enforce the principle of least privilege~\cite{gudka2015clean}. However, all these approaches rely on an initial set of developer-provided annotations~\cite{gudka2015clean}.

In this work, we aim to fill the gap and propose \underline{C}ode \underline{A}nnotation \underline{L}ogic (\sys) -- the tool that can automatically identify security sensitive code that needs TEE protection. Our tool is the first of its kind, it is designed to automatically analyze a given code base and determine which functions are security-sensitive and should be relocated to the TEE, as well as which can safely remain in the main execution environment. By streamlining this process, \sys empowers developers to optimize their applications for TEE usage, minimizing the TCB and enhancing overall security without the burden of labor-intensive manual code analysis.

\vspace{0.2cm}
\noindent\textbf{Contributions:} In more details, we make the following contributions: 

\begin{itemize} 
    \item \textbf{Security-sensitive code notion and dataset.} 
    We introduce a definition of security-sensitive code, based on the use of cryptographic functions and the confidentiality and integrity of their inputs. Using this foundational notion, we constructed a comprehensive dataset of open-source projects that employ cryptographic functions, a task that required substantial expert knowledge and intensive manual annotation, totaling approximately $5$ person-months. Our dataset, containing over $3130$ samples, fills a critical gap in automated TEE integration research. In support of open science, we will release this dataset publicly to further the development of tools and methods in secure code analysis.

    \item \textbf{Data preparation pipeline and CAL model.} 
    We developed a graph-based approach to transform code into a feature-rich graph representation. Our pipeline introduces unique capabilities, such as constructing multiple node features like syntactic node types, long-range dependencies via Node2Vec~\cite{node2vec}, semantic textual attributes (e.g., data types), and structural graph metrics. These novel features enable a more accurate and context-sensitive representation of security-sensitive code. This prepared data is processed by our custom-built CALGNN model, which leverages GNNs to detect security-sensitive nodes within the graph. These nodes are then mapped back to functions in the original codebase, providing line-accurate code annotations.
    
    \item \textbf{Evaluation and end-to-end use case.} \sys demonstrates robust performance, achieving a high recall of 86.05\%, an F1 score of 81.56\%, minimizing both false negatives and false positives. \sys achieved an identification rate of 91.59\% for correctly identifying security-sensitive functions. 
    Further validation through a case study involving a Bitcoin utility tool showed that \sys effectively identified security-sensitive code regions with high probability while correctly assigning low probabilities to non-sensitive ones. This practical demonstration highlights \sysNoSpace’s potential to support developers in creating secure applications that optimally use the TEE capabilities.

\end{itemize}

This pioneering dataset and graph-based model together create a robust framework for automating TEE integration, paving the way for efficient, secure application development with a minimized TCB. By accurately identifying the truly security-sensitive components, our approach ensures that unnecessary code is excluded from the TEE, reducing potential vulnerabilities within the trusted environment. Through the use of \sys framework, developers can leverage TEEs to their fullest potential, optimizing both performance and security while mitigating risks associated with overly broad code inclusion in trusted environments.

\vspace{0.2cm}

\noindent\textbf{Outline.} The rest of the paper goes as follows. Section~\ref{sec:background} provides the foundational background on key concepts. Section~\ref{sec:problem} defines the problem of security-sensitive code identification, highlighting the challenges involved. In Section~\ref{sec: approach}, we present our \sys approach in details. Section~\ref{sec:Evaluation} discusses the evaluation of our method,  showcasing its effectiveness through various metrics and case studies. Section~\ref{sec:related_work} reviews the related literature. Finally, Section~\ref{sec:conclusion} concludes the paper.

\section{Background}
\label{sec:background}
This section introduces the foundational concepts essential for understanding our approach, demonstrating how we leverage graph-based learning to automatically identify security-sensitive code.

\subsection{Graph Neural Networks}
\label{sec:gnn}
Graph neural networks (GNNs) are a class of neural networks designed particularly to enable effective learning on graph-structured data. 
GNNs work by representation learning of graphs capturing the complex relationships and global and local dependencies between nodes. 
This principle gives them the ability to make more sensible predictions about entities in a graph or graph itself. 
To achieve this, GNNs transform nodes and edges into numerical vector representations, known as embeddings, which provide a consistent way to encode the properties and relationships of each graph element. 
To better understand this process, it is helpful to explore the graph representation pipeline used by GNNs and its key elements. 

\vspace{0.2cm}\noindent\textbf{Key Elements of Graph Representation Pipeline:} 
The key elements of GNN's representation pipeline are \emph{node} and \emph{edge embeddings}, \emph{graph structure} and \emph{message passing} and \emph{classification layer}. 

\emph{Node and edge embeddings} represent the initial and possibly learned features of nodes and edges, providing the raw inputs for message passing in the form of a continuous vector. Node embeddings capture the node’s features and interactions, while edge embeddings capture the nature of the relationship between connected nodes. 
\emph{Graph structure} represents the connectivity between nodes, typically using an adjacency matrix or an edge list, which guides the flow of information throughout the network. This structure allows operations to consider local relationships (immediate node connections and interactions) and global relationships (patterns and dependencies across the entire graph).

\emph{Message passing} defines the interaction and aggregation functions that combine information across nodes and edges based on the graph structure, iteratively updating the embeddings. The nodes aggregate information from their neighbors, updating their own representations based on received messages. This aggregation is done iteratively in layers, allowing each node to gather information not only from its direct neighbors but also from neighbors several “hops” away. For example, after two layers, each node’s embedding will contain information from nodes up to two edges away. This helps each node build a representation that captures the context of its neighborhood.

\emph{Classification layer} After message passing, each node’s final embedding is fed through a classification layer, typically a softmax layer that outputs the probability of each possible class label for that node.

Together, these key elements form the graph representation learning components of a GNN, working together to learn meaningful representations of nodes, edges, and the entire graph.

\vspace{0.2cm}\noindent\textbf{GNN Learning Mechanisms:} 
Various GNN learning mechanisms have been developed based on different ways of aggregating node information and capturing neighborhood relationships. In our work, we explore two mechanisms that we found particularly effective for identifying security-sensitive code: 

\textbf{Gated GNN \cite{gatedGNN}}: Gated GNNs, instead of simply updating node embeddings based on passed messages, apply a gating mechanism inspired by recurrent neural networks. The gating mechanism decides how much of the new information to retain versus how much of the previous state to carry over. This allows the network to handle long-term dependencies in the graph structure by selectively “remembering” or “forgetting” information across multiple message-passing rounds. These gated updates are applied iteratively over several message-passing steps (often called “layers” in GNNs). Each iteration allows nodes to incorporate information from increasingly distant neighbors, gradually building a more holistic representation of the graph.

\textbf{GAT \cite{gatv2}}: Graph Attention Networks (GATs) introduce the concept of attention mechanisms to GNNs, allowing nodes to weigh the importance of their neighbors during message passing. The attention mechanism allows GATs to adaptively assign importance to each neighbor. This approach is beneficial in code graphs where certain connections are more relevant than others. 

GATs and Gated GNNs are powerful extensions of basic GNNs, with GATs suited to tasks needing adaptive neighbor selection and Gated GNNs suited to scenarios with long-range dependencies.

\vspace{0.2cm}\noindent\textbf{Node Classification:} Node classification refers to the task of assigning a label or class to each node in a graph based on its features, as well as the structural information of the graph. This is one of the core applications of GNNs, often seen in social networks, citation networks, and more. 
In the context of program analysis, node classification provides a promising approach for identifying security-sensitive code by treating code components — such as functions, instructions, or program variables — as nodes within a graph. The relationships between these nodes, represented by edges, can capture control flow, data flow, or dependency information within the code. Using GNNs for node classification enables the model to learn both the distinct features of each code component and the context provided by its interactions with other components in the program.

\subsection{Code Property Graph}
\label{sec:CPG}
Code Property Graphs (CPGs)~\cite{CPG} provide a versatile and expressive representation of program code by merging multiple types of code relationships — control flow, data flow, and syntactic structure — into a single, unified graph. This rich, multi-faceted structure makes CPGs particularly relevant for GNNs, enabling GNNs to leverage comprehensive program information in a single framework. 

Specifically, a CPG integrates three key code representations: the Abstract Syntax Tree (AST)~\cite{ASTs}, the Control Flow Graph (CFG)~\cite{CFG}, and the Program Dependence Graph (PDG) ~\cite{PDG}. 
The AST represents the syntactic structure of the code, capturing hierarchical relationships between different code elements.
The CFG describes the potential execution paths within the program, while the PDG captures control and data dependencies. 
By merging these representations, CPGs enable the analysis of both the structural and behavioral aspects of the program in a single unified graph.

The CPGs are encompassed with nodes and edges with meaningful connections. 
Nodes in a CPG represent different program elements such as variables, functions, expressions, and control constructs. 
These nodes are enriched with detailed attributes, including types, identifiers, and relationships, which allow them to encode rich syntactic and semantic information about the source code.
For instance, node types such as \texttt{METHOD} (which represents function definitions), \texttt{TYPE\_DECL} (which defines type declarations), \texttt{CONTROL\_STRUCTURE} (which represents constructs like loops and conditionals), and \texttt{CALL} (which represents function or method invocations) are just a few examples. 
Together, these different node types help distinguish various components of the program.

These nodes are interconnected by the edges in the CPG, which are relationships such as control dependencies comprised of CFGs, data dependencies of PDGs, and structural relationships from the AST. This integration enables a comprehensive code analysis that captures diverse aspects of program behavior and structure, allowing for a unified view that links control, data, and syntactic information within the program.

Overall, CPGs provide a unified representation that captures both syntactic and semantic code patterns in subtle detail, making them an ideal underlying graph structure for machine learning models. When combined with GNNs, CPGs enable models to effectively learn complex relationships within code, making this approach particularly well-suited for identifying security-sensitive code. 

\section{Problem Statement}
\label{sec:problem}

In today’s security-critical applications, precisely separating code into security-sensitive and non-sensitive components is paramount for efficient and secure integration with TEEs. This partitioning minimizes the TCB, reducing the risk of vulnerabilities within the trusted environment. However, achieving this split is not straightforward, as it requires a clear definition of "security-sensitive code," which remains challenging due to the varied nature of application contexts and the diverse threat landscape.

This section addresses the foundational aspects of this challenge, beginning with a definition of security-sensitive code, which underpins our approach (cf.~\Cref{sec:problem:notion}). We then describe the attacker model to establish the assumptions and capabilities of potential adversaries (cf.~\Cref{sec:problem:attacker_model}). Finally, we highlight the unique challenges in identifying security-sensitive components in code using GNNs, which demand more sophisticated analysis than typically performed other analysis tasks, such as vulnerability detection (cf.~\Cref{sec:problem:challenges}). Together, these components frame the complex problem of security-aware code partitioning and set the stage for our contributions toward a practical solution.

\subsection{Security Sensitive Code}
\label{sec:problem:notion}
In this paper, we use a notion of security-sensitive code that is defined based on the use of cryptographic operations. 
Cryptographic functions lie at the heart of secure software design, safeguarding the confidentiality, integrity, and authenticity of data. By their very nature, these functions are used to process or protect sensitive information, whether it’s user credentials, financial transactions, or proprietary data. Cryptographic operations, such as encryption and decryption, digital signatures, and hashing, directly influence an application’s ability to resist unauthorized access and manipulation, which makes them critical points of interest for attackers.

The sensitivity of cryptographic functions stems from their role in securing data against numerous threat vectors. Compromising these operations exposes sensitive data to potential misuse, undermines the trustworthiness of the application, and can even lead to broader system breaches. For instance, improper handling of encryption keys could allow attackers to decrypt confidential information, while a compromised hashing function might enable the tampering of data without detection. Furthermore, cryptographic operations are often designed with strict requirements for secrecy and correctness, as even minor vulnerabilities in these functions can yield catastrophic security consequences, such as leaking private data or permitting unauthorized access. Hence, in software projects that integrate TEEs, cryptographic functions must always be executed within the TEE to ensure the trustworthiness of their outputs. 

In addition to cryptographic functions, the inputs provided to these functions—such as data to be encrypted, passwords, or cryptographic keys—are equally security-critical and require protection. These inputs represent sensitive information that, if exposed or altered, could undermine the effectiveness of cryptographic operations, leading to potential breaches in confidentiality and data integrity. Protecting these inputs within a TEE ensures that they remain isolated from untrusted parts of the system, safeguarding them from unauthorized access or tampering.

From a dataflow perspective, cryptographic functions can be seen as the “sink” of security-sensitive dataflows, where sensitive inputs converge and undergo transformation to secure their confidentiality or integrity. Under this model, our notion of security-sensitive code extends beyond the cryptographic operations themselves to encompass the entire dataflow, including all components that interact with or transmit sensitive inputs to these functions. By capturing and securing all elements in this path, we aim to provide comprehensive protection to any information that flows into cryptographic functions, be it security-sensitive data, passwords, or cryptographic keys.

\subsection{Threat Model}
\label{sec:problem:attacker_model}
In the context of identifying security-sensitive code, we consider a scenario in which developers must safeguard critical components of their applications against sophisticated attackers. This threat model specifically targets a powerful adversary with kernel-level access to the system, who seeks to exploit software vulnerabilities in order to compromise the integrity or confidentiality of sensitive data and code. By leveraging control over the system, the attacker aims to breach isolated components or extract sensitive information.

However, we assume that the adversary cannot compromise code running under Trusted Execution Environment (TEE) protection, nor can they interfere with the operation of \sys itself. This assumption is reasonable, as \sys is applied prior to code deployment on the system, during the code development process which is not affected by adversaries.

\subsection{Challenges}
\label{sec:problem:challenges}
In developing a reliable framework for identifying security-sensitive code, several challenges emerge that must be addressed to enhance the effectiveness of automated tools and methodologies. These challenges stem from the inherent complexities of software systems, the nuances of data flows, and the limitations of current analysis techniques. Below, we outline the primary challenges encountered in this endeavor, each highlighting critical aspects of the problem.

\vspace{0.2cm}\noindent\textbf{CH I - Complexity of the  Security Sensitive Code Notion:}\label{ch:challenge1} 
The first challenge lies in accurately identifying sources of security-sensitive data flows.
Each function must be evaluated to determine if its execution could impact the confidentiality and integrity of sensitive data used in cryptographic operations. Assessing whether each function could serve as a valid entry point for a security-sensitive data flow is crucial to ensure that placing a TEE boundary around it does not compromise the security of the inputs involved. This requires analyzing all possible function invocations within the application to determine if any could violate security guarantees when isolated within a TEE.

Static analysis faces limitations due to its lack of runtime context, meaning it cannot account for dynamic factors like user inputs or environmental variations, which makes modeling complex control flow and data dependencies challenging. Dynamic execution, on the other hand, struggles with incomplete coverage, as it is practically impossible to exercise every possible code path, particularly in large applications. As a result, accurately identifying security-sensitive functions remains a significant challenge.

\vspace{0.2cm}\noindent\textbf{CH II - Dataset:}\label{ch:challenge2} 
The second major challenge is the lack of existing datasets to support the training and validation of models for identifying security-sensitive code. Reliable identification tools require a substantial volume of labeled data, but no publicly available datasets categorize code by security sensitivity, particularly for cryptographic operations and TEE integration. Manually creating such datasets is labor-intensive and requires expert knowledge to accurately identify and label security-sensitive elements, such as specific code sections, functions, or data flows that need protection to maintain security guarantees.

\vspace{0.2cm}\noindent\textbf{CH III - Long-Range Security-Sensitive Dataflows:} \label{ch:challenge3} 
Security-sensitive data flows frequently span long chains of function calls, creating complex, long-range dependencies. Unlike typical vulnerability detection tasks that focus on localized interactions, identifying security-sensitive code requires tracking data dependencies across multiple functions within the application. This complexity demands advanced analysis techniques that can capture and model these extended relationships in the codebase. Effectively managing these long-range dependencies is crucial for accurately identifying sensitive data paths and ensuring that all relevant sections of the code are properly secured.

\vspace{0.2cm}\noindent\textbf{CH IV - Limitations of modern GNN-based tools:} \label{ch:challenge4} Current GNN-based vulnerability detection methods often rely on simplistic features like syntactic node types, which fail to capture the complexities involved in identifying security-sensitive code. Existing tools \cite{cao2021bgnn4vd, Devign, DeepWukong, VulChecker, hin2022linevd, PAVUDI, siow2022learning, CPVD, UnifiedCPG, zhang2023dshgt} treat calls to non-critical and cryptographic library functions similarly in their graph representation, challenging effective differentiation. 

Additionally, the inability to distinguish between different function calls complicates data flow analysis within the application, hindering a clear understanding of how sensitive data is managed. 
These limitations pose a significant challenge in identifying which functions require special security considerations.

\vspace{0.2cm}\noindent\textbf{Summary:} Addressing the challenges of identifying security-sensitive code requires a comprehensive understanding of data flow complexities, robust training datasets, techniques for capturing long-range dependencies in function calls, and detailed feature representations that differentiate sensitive functions from nonsensitive ones. 
Overcoming these obstacles is essential for developing effective automated tools that can enhance application security while optimizing the integration of TEE.

\section{Code Annotation Logic}
\label{sec: approach}

\begin{figure}[t]
    \centering
    \includegraphics[width=1.1\columnwidth]{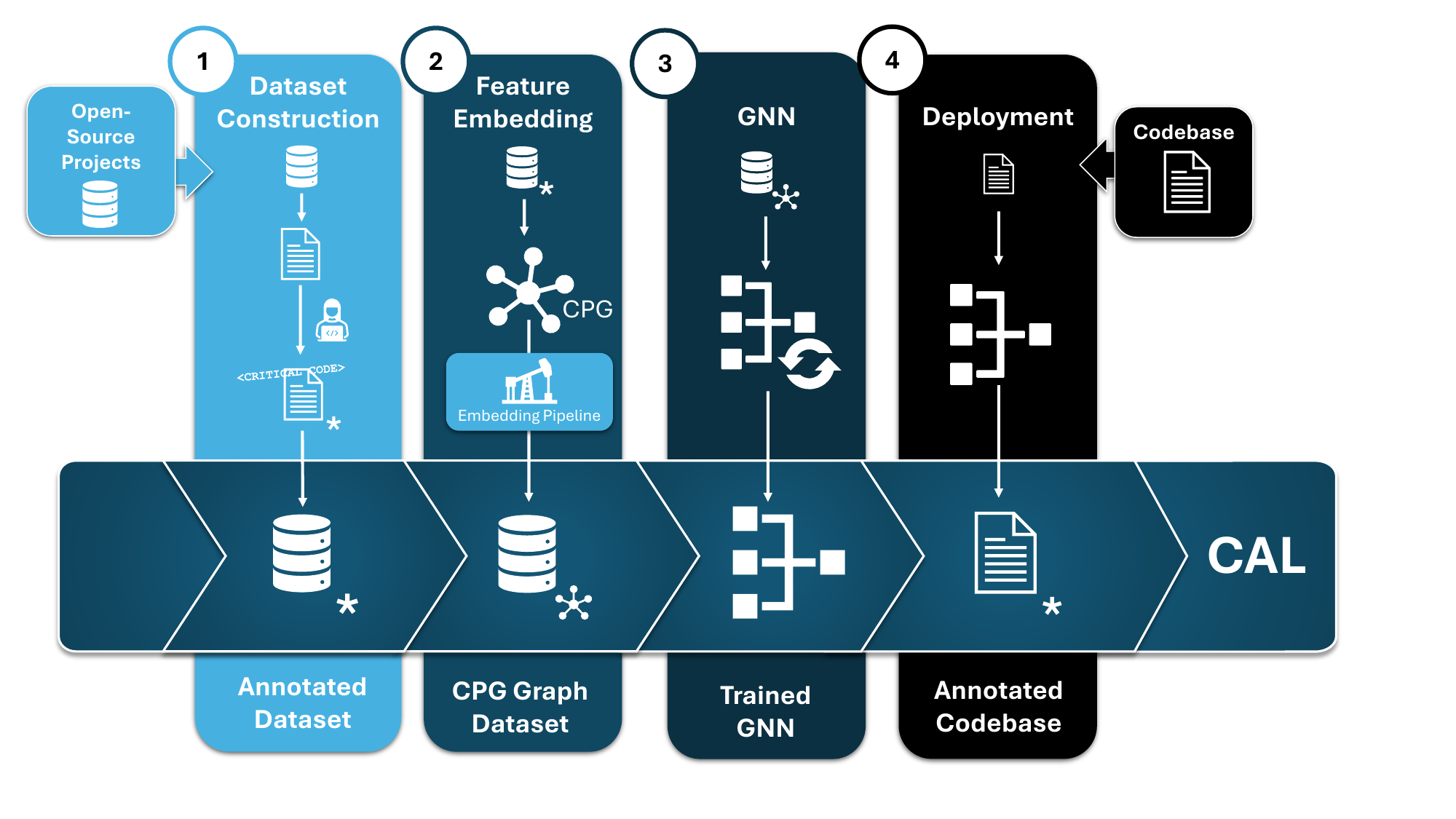} 
    \caption{\sys Phases Overview}
    \label{fig:cal_phases_overview}
\end{figure}

To tackle the identified challenges in automating the identification of security-sensitive code, we developed the Code Annotation Logic (CAL) tool, which leverages advanced techniques in graph-based analysis and machine learning.

To address the first challenge \CHoneRef~identifying security-sensitive functions according to our definition of security sensitivity, we employed GNNs. GNNs have demonstrated efficacy in analyzing and processing complex software systems, making them well-suited for analyzing non-trivial relationships between functions and their data flows.

In response to the second challenge \CHtwoRef, the absence of suitable training datasets, we constructed the first-ever dataset of its kind, consisting of $1070$ manually labeled open-source software projects. This dataset serves as a crucial resource for training and evaluating our model, providing a rich foundation for understanding the characteristics of security-sensitive code.

To solve the third challenge of capturing long-range security-sensitive data flows \CHthreeRef, we introduced more advanced feature engineering techniques. These techniques allow our model to effectively capture and model the relationships among functions that span extensive call chains, thereby enhancing its ability to recognize security-sensitive interactions across the codebase.

Finally, to tackle the fourth challenge related to distinguishing between critical and non-critical function calls \CHfourRef, we further refined our feature engineering techniques. By integrating textual information from the source code into our graph representations, we enhance the model's ability to differentiate between ordinary functions and those that are security-sensitive. This differentiation is vital for ensuring accurate annotations and robust security assessments.

\noindent\textbf{High-level Overview:} \sys is a graph-based tool that transforms the source code into a CPG and applies a custom feature engineering pipeline specifically tailored to the challenges of detecting security-sensitive code. The resulting graphs are then processed by a custom-built and trained CALGNN, which performs node classification by assigning a probability to each node, indicating whether it is security sensitive. 

In our approach, we define the boundaries of security-sensitive code at the function level, considering entire functions as security-sensitive rather than isolating specific lines or segments within them. This design choice is driven by practical considerations, as isolating parts of a function for TEE integration could lead to frequent switches between the TEE and non-TEE environments, resulting in significant overhead. Moreover, when sensitive data flows through multiple functional layers, attempting to secure only fragments of a function increases complexity and the risk of inadvertently exposing sensitive information to unsecured code paths. By focusing on entire functions, we ensure a more efficient and secure handling of security-sensitive operations.

As illustrated in \Cref{fig:cal_phases_overview}, the development process of \sys involved four main phases. In Phase 1, we defined a labeling process and guidelines and manually processed hundreds of C projects to create a dataset that matched our notion of security-sensitive code~\Cref{sec:Phase 1}. In Phase 2, we designed a feature embedding process to address the specific challenges of security-sensitive code identification, focusing on incorporating long-range and textual features to enable the detection of cryptographic library usage~\Cref{sec:Phase 2}. In Phase 3, we custom-built our GNN architecture for the problem of security-sensitive code identification and trained it on the dataset. The GNN, therefore, learns the patterns of security-sensitive code and can then later apply this knowledge to unseen code during inference~\Cref{sec:Phase 3}. Finally, in Phase 4, we deployed the trained GNN along with the feature engineering pipeline, providing developers with an end-to-end annotation system that processes a codebase and outputs a list of the security-sensitive functions that should be moved to the TEE~\Cref{sec:Phase 4}. 

\subsection{Phase 1 - Dataset Construction}
\label{sec:Phase 1}
As outlined in \Cref{sec:problem:challenges}, the successful implementation of a machine learning-based approach for modeling and automatically detecting security-sensitive functions in previously unseen projects necessitates a comprehensive dataset wherein these functions are explicitly labeled. To address \CHtwoRef, we created such a dataset by collecting relevant open-source projects from GitHub~\cite{github} that utilize cryptographic libraries. We subsequently implemented a structured manual labeling process designed to ensure that the annotations accurately reflect our definition of security-sensitive code.

\begin{enumerate}[noitemsep,nolistsep]
    \item \textbf{Filtering Ineligible Projects:} Our process commences with the exclusion of projects that lack a diverse mix of both security-sensitive and non-security-sensitive functions. 
    \item \textbf{Identification of Cryptographic Functions:} We then systematically identify all instances of calls to cryptographic libraries or custom cryptographic implementations within the selected projects. These functions serve as initial indicators of security-sensitive code. 
    \item \textbf{Establishing an Initial Set:} Not all calls to cryptographic libraries are security-sensitive, such as those related to error-handling functions. These non-sensitive calls are excluded, focusing instead on functions that implement cryptographic operations, which form the initial set of security-sensitive code to be further refined in subsequent analytical phases. 
    \item \textbf{Snowballing:} We engage in a recursive exploration of the functions that either invoke or are invoked by these security-sensitive functions. \emph{Backward Snowballing} examines the functions called by security-sensitive functions, as they are often security-sensitive themselves. \emph{Forward Snowballing} analyzes the functions that invoke security-sensitive functions, evaluating how they handle sensitive data to ascertain if they warrant designation as security-sensitive.
    \item \textbf{Global Variable Analysis:} In certain instances, global variables may store security-sensitive data. We systematically track the utilization of such variables and, subsequently, mark any functions that interact with them as security-sensitive when appropriate. 
\end{enumerate} 

\begin{figure*}[t]
    \centering
    \includegraphics[width=0.95\textwidth]{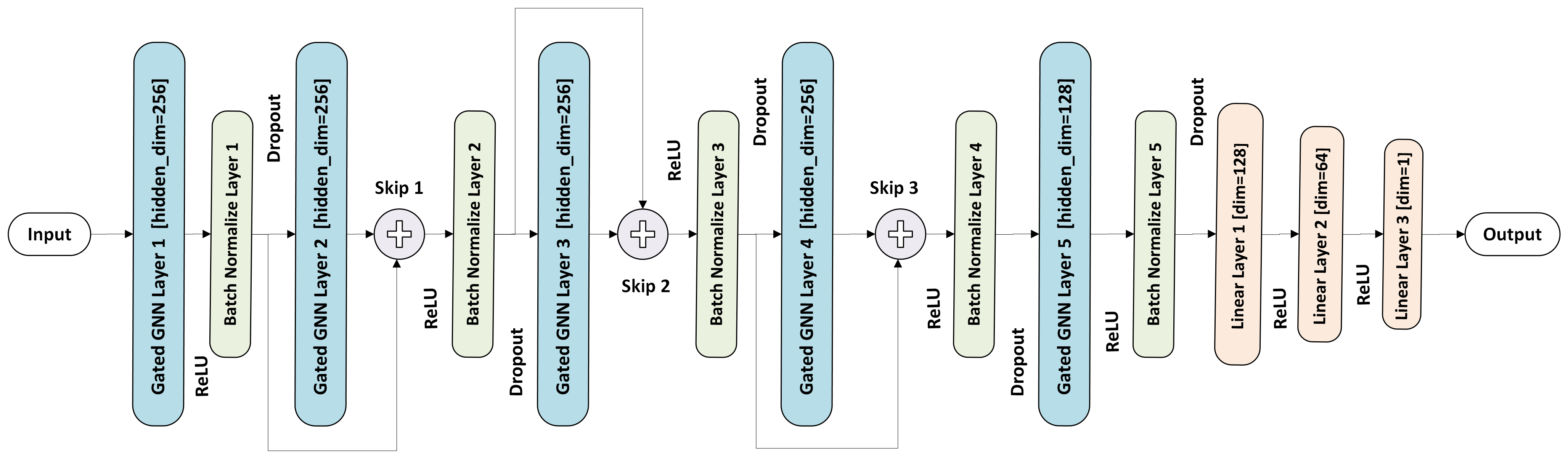}  
    \caption{CALGNN Architecture Schematic}
    \label{fig:gnn architecture schematic}
    \vspace{-1.5em}
\end{figure*}

\subsection{Phase 2 - Feature Embedding}
\label{sec:Phase 2}
To create an effective representation of the code for identifying security-sensitive code, we developed a comprehensive pipeline to convert source code into graph representations.
Specifically, we utilized Joern~\cite{joern}, a well-established tool for generating CPGs, which provide a unified view of the structural and semantic aspects of the code.
These code representations serve as the foundation for our code graphs, which are then converted into dense vector representations to be fed into GNNs for effective learning and precise pattern identification.
As stated in~\Cref{sec:gnn}, this encoding step is an essential requirement for GNNs to process code graphs. 
However, the existing approaches that leverage these code graphs, for instance, in vulnerability detection, often rely solely on the node embeddings derived from the syntactic elements of the graphs, such as variables, functions, expressions, and control constructs, which limits their ability to capture deeper, context-specific information needed for secure-sensitive code identification.
To address all these challenges as defined in~\Cref{sec:problem:challenges}, we built a novel feature construction pipeline and embedded all this rich information into node and edge features of the code graphs. 
These features, such as long-range dependencies, textual indicators of cryptographic usage, and graph metrics that provide structural insights, augment the syntactic node embeddings to offer a comprehensive representation of the source code. In detail:

\vspace{0.2cm} \noindent \textbf{Long-Range Dependencies:} 
To address \CHthreeRef, we employed Node2Vec~\cite{node2vec} as a method to augment the node features with additional contextual information derived from the overall graph structure.
Node2Vec algorithm performs random walks to capture both local and global graph structural relationships for each node. 
We carefully tuned algorithm parameters to emphasize long-range relationships over neighboring nodes.
This technique and tuning are particularly beneficial in the context of software code, where functions may interact in complex and extensive call chains. 
By capturing these long-range dependencies, Node2Vec enhances the ability of the model to identify security-sensitive nodes that are influenced by distant parts of the code. 
This is crucial in scenarios where the security impact of a function is not immediately apparent from its direct neighbors but is revealed through indirect interactions across the codebase. 
Moreover, these pre-computed long-range features complement the GNN layers in the CALGNN model, which primarily focuses on capturing local neighborhood information.
Together, this provides a holistic understanding of both local and global dependencies, ultimately improving the accuracy of identifying security-sensitive code.

\vspace{0.2cm} \noindent\textbf{Textual Features:} 
To address \CHfourRef, we enhanced the node embeddings by incorporating additional textual features that provide explicit and valuable semantic context from the source code that helps treating cryptographic library and non-critical functions differently. 
Unlike basic syntactic node types, which broadly classify nodes, these features capture specific attributes such as data types, offering more semantic context that can be particularly useful for identifying security-sensitive code elements. 
Therefore, we specifically utilized the ‘datatype’ attribute from Joern's CPG for each node to extract this useful information.
For instance, attributes like \texttt{TYPE\_FULL\_NAME} in CPG's provide details about types of variables (e.g., \texttt{EVP\_PKEY}, \texttt{char}, \texttt{int}). \texttt{EVP\_PKEY} represents a cryptographic key (public, private, or both) in OpenSSL, making it one of the key indicators of cryptographic operations. 
Capturing all these details is crucial for understanding the specific role of each node, particularly in security-sensitive cryptographic contexts.

To convert these textual attributes into useful node features, we collected relevant content from the nodes in each code graph and trained a Word2Vec~\cite{word2vec} model to encode them into dense vector representations. 
Word2Vec, a well-established technique for generating word embeddings, ensures that semantically similar values—such as datatypes—are mapped closely in the feature space.
These vector representations help the \sys model distinguish between different nodes based on their specific roles.

\vspace{0.2cm} \noindent\textbf{Graph Metrics:}  
Finally, we complement the other node embeddings with graph-theoretic metrics that provide significant structural insights. 
Specifically, we explored three key indicators such as node degree, closeness centrality, and betweenness centrality, that can provide distant information about the code graphs, which can directly impact the identification of secure sensitive code.

\noindent\textbf{Node degree \cite{newman2010networks}}\label{node_degree} measures the number of direct connections a node has, reflecting its immediate connectivity and importance within the code graph. For instance, a function definition node with a high degree may represent a function that is frequently called or interacts with multiple functions, potentially elevating its significance in the context of security sensitivity.

\noindent\textbf{Closeness centrality \cite{closenesscentrality}}\label{closeness_Centrality} assesses how quickly a node can access other nodes in the graph, indicating its overall reachability. 
Nodes with high closeness centrality can be crucial for understanding data flow throughout the code, as they could potentially propagate secure sensitivity across multiple components. 

\noindent\textbf{Betweenness centrality \cite{betweenness_Centrality}}\label{betweenness_centrality} identifies nodes that act as bridges between other nodes in the graph. 
For instance, a function node with high betweenness centrality may control critical data pathways, making it a significant point for ensuring security-sensitive operations.

Since these metric values vary for each node in the code graph, they are useful for distinguishing different node roles, and our experiments showed that incorporating them improved the CALGNN model's ability to identify secure-sensitive code.

\subsection{Phase 3 - Graph Neural Network}
\label{sec:Phase 3} As discussed in~\Cref{sec:problem:challenges} \CHoneRef, static methods are inadequate for detecting security-sensitive functions.  To overcome this limitation, we turn to machine learning, specifically employing a GNN to automate the identification of security-sensitive functions in arbitrary C projects. 

Our tool, \sys, utilizes a custom-built CALGNN model to analyze the graphs generated by our feature engineering pipeline.

\begin{figure*}[t]
    \centering
    \includegraphics[width=0.9\textwidth]{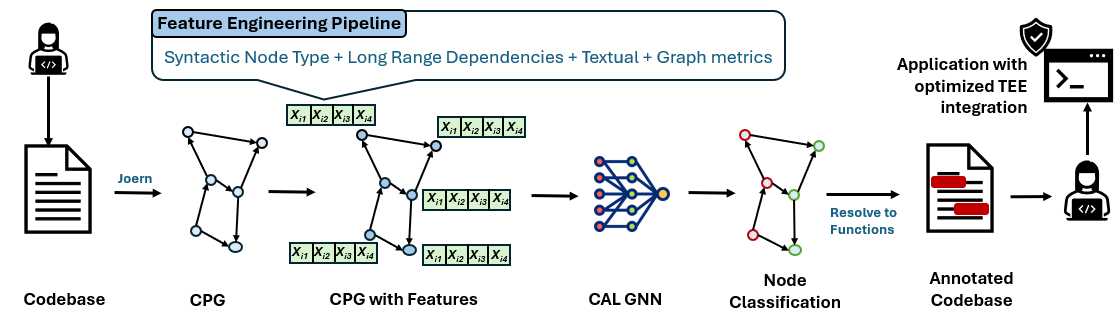} 
    \caption{\sys Deployment Workflow}
    \label{fig:cal_deployed}
    \vspace{-1.5em}
\end{figure*}

\subsubsection{CALGNN Model}
The CALGNN model is designed to accurately identify security-sensitive nodes within CPGs, effectively distinguishing between secure and non-secure sensitive patterns in source code. 

The model leverages advanced GNN learning mechanisms~\cite{gatedGNN, gatv2} to learn complex relationships within the code graph and employs effective training strategies to enhance robustness, generalizability, and classification performance.

\vspace{0.2cm} \noindent{\textbf{GNN Architecture:}}
As illustrated in \Cref{fig:gnn architecture schematic}, the optimal GNN architecture of \sys is a deeper and complex design that consists of multiple Gated GNN layers \cite{gatedGNN}, multiple skip connections and a sequence of \textit{Linear Layers}  to effectively capture the complex relationships inherent in code structures. 
Gated GNN layers are used to learn rich representations of node-level features, allowing the model to effectively comprehend both structural and contextual information present in the code graph. 
In \Cref{fig:gnn architecture schematic}, the term hidden\_dim represents the hidden dimension used in each layer. The model starts with a larger size of $256$ to provide sufficient capacity for capturing complex interactions and dependencies between nodes. This larger dimension allows the model to learn intricate features effectively. As we progress through the layers, it is reduced to $128$, which helps distill the learned features, reduce computational costs, and emphasize the most significant patterns for accurate final predictions. 
With an increased depth of architecture, the model is capable of better representation learning, capturing complex code-level relationships that shall help distinguish secure-sensitive nodes from non-sensitive ones.
Skip connections are added between successive GNN layers to ease gradient flow and mitigate vanishing gradient issues, which are helpful in deeper networks. 
Each GNN layer is followed by ReLU activation, Batch Normalization, and Dropout to ensure stable training and prevent overfitting.

Finally, classification is achieved through a series of linear layers that progressively reduce dimensionality, ensuring that the learned features are effectively combined to produce accurate predictions of secure sensitive nodes.
To complete the classification, the model's final linear layer is equipped with a sigmoid activation function, which produces a probability score ranging from $0$ to $1$ for each node. 
A higher score signifies an increased likelihood that the node is part of a security-sensitive function. 
Nodes scoring above a threshold of $0.5$ are labeled as security-sensitive and mapped back to their corresponding lines of code using custom functions as described in~\Cref{sec:appendix}. 
If any node within a line of code is classified as security-sensitive, the line itself is flagged as security-sensitive. Additionally, if any line within a function is classified as security-sensitive, the entire function is flagged, ensuring comprehensive coverage.

\subsection{Phase 4 - Deployment}
\label{sec:Phase 4}
In the final phase, we deploy the trained GNN model alongside Joern~\cite{joern} and the feature embedding pipeline to establish a comprehensive system that streamlines the integration of security-sensitive code identification into the software development process. 
The workflow, illustrated in \Cref{fig:cal_deployed}, provides a seamless, end-to-end approach for identifying and isolating secure-sensitive components within a codebase. 
In this deployment workflow, developers begin by providing the codebase in text format.
Joern~\cite{joern} is used to generate CPGs, which form the basis for our analysis. 
These generated CPGs undergo further enhancement via a feature engineering pipeline, which enriches the graph with multiple types of node features: syntactic node type, long-range dependencies, textual information, and graph metrics such as node degree and centralities as defined in~\Cref{sec:Phase 2}.
As illustrated in \Cref{fig:cal_deployed}, each node feature type is denoted using $X_{i1}, X_{i2}, X_{i3}, X_{i4}$, where these symbols represent different feature sets corresponding to each node in the graph. These node features are concatenated to form a rich, dense vector representation for each node, which enables the CALGNN model to learn and identify security-sensitive nodes more effectively.

After feature enrichment, the CPGs with these enhanced features are fed into the trained CALGNN model for inference. 
During this step, each node is classified as either security-sensitive or non-sensitive based on the learned representations. 
As depicted in~\Cref{fig:cal_deployed}, nodes outlined in red represent security-sensitive nodes, while those outlined in green represent non-sensitive nodes.
The nodes labeled are then mapped back to their respective line numbers as described in~\Cref{sec:appendix}. 
Using these identified line numbers, we resolve their corresponding function names within the original source code. 
This capability enables developers to effectively partition functions into TEEs, thereby optimizing TEE integration and significantly reducing the attack surface of their applications.
Through this deployment phase, \sys offers a user-friendly interface for developers, enabling them to enhance the security of their codebases seamlessly. By automating the detection of security-sensitive functions and facilitating their secure execution, our tool not only promotes best practices in software security but also contributes to the creation of more resilient applications.

\section{Evaluation}
\label{sec:Evaluation}
 
This section evaluates \sys in identifying security-sensitive information at the node, line, and function levels, emphasizing high recall and efficiency. We describe the dataset, experimental setup, training strategy, and evaluation metrics. The evaluation also covers varying software project sizes and different software contexts. Finally, we present a case study assessing \sys's performance on a real-world project involving cryptographic usage. 
A comprehensive ablation study detailing the impact of each component of \sys is provided in~\Cref{sec:appendix}.

\subsection{Dataset}
To create a robust foundation for training and evaluating the \sys model, we collected and constructed a first-of-its-kind, line- and function-level dataset following the guideline process outlined in \Cref{sec:Phase 1}. 
To provide key insights into the dataset, this section covers aspects such as its coverage, and the effort involved while also detailing the de-duplication techniques used to ensure a clean and high-quality dataset. 
The details about the data split and size are mentioned in~\Cref{sec:appendix}.

\vspace{0.2cm} \noindent\textbf{Coverage:} We have collected the source code of $1070$ open-source software projects written in $C$ from GitHub, each utilizing the OpenSSL~\cite{opensslweb} library. 
Specifically, these projects cover various categories, including cryptographic algorithms and implementations, OpenSSL integration and extensions, security tools and libraries, networking, and secure communication, as well as application development with platform-specific implementations. 
The projects, therefore, interact with the OpenSSL in diverse ways, which ensures that \sys trained on our dataset can generalize effectively and be applied to arbitrary projects using the library. 
The largest project in our dataset contains $13,428$ lines of code, while the median size across all projects is $492$ lines of code. 

OpenSSL was selected as the underlying cryptographic library for our dataset as it is one of the most widely used libraries for secure communications, with extensive deployment across a broad spectrum of applications and platforms. This widespread usage provides a rich and varied dataset, capturing numerous interaction patterns and edge cases with a well-established cryptographic standard. By focusing on OpenSSL, we ensure that the trained model is relevant to a significant proportion of real-world applications.

\vspace{0.2cm}\noindent\textbf{Deduplication:} To ensure a unique dataset, prevent bias, and maintain reliable performance metrics, we removed duplicate samples.
A graph-based deduplication technique was used, where each CPG was hashed using the SHA-256 algorithm, capturing node and edge attributes. 
Identical hashes indicated duplicate graphs, allowing us to effectively eliminate approximately 17.06\% of redundant samples.

\vspace{0.2cm}\noindent\textbf{Effort:} The creation of the dataset required substantial manual expert annotation effort, carried out by a dedicated team of students and researchers, with guidance from industry experts. 
More than 1100 hours were spent labeling security sensitive code sections across all projects, ensuring accuracy and consistency throughout the dataset.

\subsection{Preliminaries}
This section presents the experimental setup, training strategy, and evaluation metrics used in our approach to assess the performance of \sys.
\subsubsection{Experiment Setting}
This subsection summarizes the hardware, data splitting strategy, optimization framework for parameter tuning, class imbalance solutions, and regularization techniques used to train and optimize the \sys{} models effectively.

\vspace{0.2cm} \noindent\textbf{Hardware:}
Training and evaluation were conducted on a Linux server equipped with an Intel(R) Xeon(R) Gold 6342 CPU (64-bit) and 251GB of RAM. Additionally, CUDA~\cite{nvidia2020cuda} and an NVIDIA A30 GPU with 24GB of memory were utilized to accelerate the training process. 

\vspace{0.2cm} \noindent\textbf{Stratified $K$ Fold Cross Validation:}
To ensure a balanced and effective training process, we used stratified $k$ fold cross-validation instead of a single train-test split. 
This method ensures that each fold maintains a proportional representation of secure-sensitive nodes, providing consistent learning while preventing overfitting. 
It also enhances the robustness and generalizability of the model by ensuring a uniform distribution of security-sensitive samples throughout the training process. 
The optimal number of folds, $K$, was determined experimentally. 

\vspace{0.2cm} \noindent\textbf{Hyperparameter Optimization:} To refine the \sys{} model's hyperparameters, we utilized Optuna \cite{akiba2019optuna}, a sophisticated hyperparameter optimization framework. 
This approach enabled efficient exploration of the hyperparameter space, optimizing parameters such as learning rate, optimizer, number of epochs, and batch size, while also ensuring early termination of less promising trials to identify the best configuration with minimal computational resources.

\vspace{0.2cm} \noindent\textbf{Class Weights:} To ensure balanced learning and mitigate bias towards the majority class, calculated class weights were used to emphasize the minority class. 
This approach helps the model focus on underrepresented secure-sensitive nodes, improving overall classification performance.

\vspace{0.2cm} \noindent\textbf{Regularization Techniques:} Lastly, to prevent overfitting of the \sys model, we employed regularization techniques such as \textit{weight decay} to penalize large weight values, \textit{early stopping} to terminate training when validation performance stagnated, and a \textit{learning rate scheduler} to adaptively adjust the learning rate, promoting smoother convergence. 

\subsubsection{Evaluation Metrics} To comprehensively assess the performance of \sys, we evaluate its ability to classify nodes within code graphs as either security-sensitive or not.
By comparing the predicted classifications with expert-labeled data curated with guidelines described in~\Cref{sec:Phase 1}, we construct a confusion matrix \cite{confusion_matrix} detailing True Positives (TP), True Negatives (TN), False Positives (FP), and False Negatives (FN). 
The following metrics, derived from the confusion matrix, are used to quantify the model's accuracy, precision, recall, and overall effectiveness.
Additionally, we introduce metrics such as Identification Rate and TCB Reduction to provide a holistic view of the model's practical impact in real-world secure environments.

\noindent$\bullet$ \textbf{Accuracy}: The proportion of correctly identified nodes, calculated as $\frac{TP + TN}{TP + FP + TN + FN}$.

\noindent$\bullet$ \textbf{Recall}: The proportion of correctly identified security-sensitive nodes, calculated as $\frac{TP}{TP + FN}$.

\noindent$\bullet$ \textbf{Precision}: The proportion of identified security-sensitive nodes that are actually security-sensitive, calculated as $\frac{TP}{TP + FP}$.

\noindent$\bullet$ \textbf{F1 Score}: A harmonic mean that combines precision and recall, calculated as $\frac{2 * (P * R)}{P + R}$. 

\noindent$\bullet$ \textbf{Identification Rate}: The proportion of actual security-sensitive lines or functions correctly identified by \sys, calculated as $\frac{\text{Identified Security-Sensitive Lines}}{\text{Total Security-Sensitive Lines}}$. 

In our context, the \sys{} model places a higher emphasis on achieving a high recall, as it is crucial to correctly identify security-sensitive nodes.
However, it is also important for \sys{} not to significantly compromise on precision, minimizing false positives.  Ultimately, the goal is for \sys{} to maintain a balanced performance, effectively identifying security-sensitive nodes while ensuring fairness and accuracy across all node classifications.
Additionally, the identification rate metric highlights \sys{}'s ability to accurately identify security-sensitive lines or functions, which is crucial for minimizing missed detections and maintaining comprehensive coverage of security-sensitive components

\subsection{\sys Performance at Different Granularity} 
In this section, we present the performance of \sys at varying levels of granularity, including node-level, line-level, and function-level evaluations, to demonstrate its ability to identify security-sensitive code effectively at each level. 
 While \sys performs well at all levels, function level evaluation is particularly important since developers typically relocate entire functions, not individual lines, to TEEs. This makes function-level identification the most practical and impactful aspect of our approach.
At the same time, the evaluation at node and line-level helps us understand the model’s ability to identify fine-grained security-sensitive patterns.
Achieving high performance at these fine-grained level is inherently more challenging compared to function-level evaluation due to the greater complexity and detail involved. 

\subsubsection{\sys Performance at Node Level}
The performance of the \sys tool, using the Gated GNN model architecture to classify nodes, is summarized in~\Cref{tab:calgnn_performance_granularity}.

The table demonstrates the system's capability to identify security-sensitive nodes within the software projects from the test set in the constructed dataset.

~\Cref{tab:calgnn_performance_granularity} shows that \sys performance at node level achieves a good recall of $86.05\%$, highlighting its effectiveness in capturing security-sensitive nodes and minimizing false negatives. 
The F1 score of $81.56\%$ reflects a balanced performance, integrating both recall and precision effectively. 
With a precision of $77.50\%$, \sys maintains an acceptable level of false positives, ensuring a reliable classification without excessive overestimation of sensitivity.
The accuracy of $76.49\%$ further supports the model's robustness by demonstrating its overall capability to correctly classify both security-sensitive and non-sensitive nodes across the entire tsst dataset.

\begin{table}[ht]
\centering
\fontsize{20}{22}\selectfont
\setlength{\tabcolsep}{15pt} 
\renewcommand{\arraystretch}{1.5} 
\resizebox{\columnwidth}{!}{
    \begin{tabular}{cccccc}
    \toprule
 \textbf{GNN Type} & \textbf{Level} & \textbf{Accuracy} & \textbf{Precision} & \textbf{Recall} & \textbf{F1-Score}\\
       \midrule
Gated GNN & Node & 76.49\% & 77.50\% & 86.05\% & 81.56\% \\ 
Gated GNN & Line & 78.68\% & 64.50\% & 95.80\% & 77.10\% \\ 
Gated GNN & Function & 79.32\% & 62.15\% & 96.23\% & 75.23\% \\ 
    \bottomrule
    \end{tabular}}
    \caption{\sys's performance at different Granularity}
    \label{tab:calgnn_performance_granularity}
    \vspace{-0.5em}
\end{table}

\subsubsection{\sys Performance at Line Level}
The performance of \sys in identifying the total number of security-sensitive lines that can be annotated automatically in various code graphs from the constructed dataset are presented in ~\Cref{tab:security_critical_lines,tab:calgnn_performance_granularity}. 
As shown in~\Cref{tab:calgnn_performance_granularity}, \sys performance at line level achieves a recall of $95.80\%$, highlighting its effectiveness in minimizing missed security-sensitive lines. 
The higher recall is due to the aggregation approach; if any node within a line is classified as security-sensitive, the entire line is flagged.
Since each line is represented in the code graph by approximately $15$ to $25$ nodes, this leads to higher recall than at the node level, but this also contributes to increased false positives at the line level, resulting in lower precision compared to the node level. 
Nevertheless, this approach ensures that no potentially sensitive line is overlooked by \sys, even if the sensitivity is subtle.

The F1 score of $77.10\%$ at the line level still indicates a balanced performance, effectively combining precision and recall.
Further evaluation uses the identification rate, a fair metric representing the proportion of actual security-sensitive lines correctly identified by \sys, providing insight into its effectiveness across different codebases.~\Cref{tab:security_critical_lines} shows that \sys achieves an overall identification rate of $90.15\%$, demonstrating its robustness in identifying security-sensitive lines of code with relatively few missed detections. 
This high identification rate underscores the \sys's capability to cover most sensitive elements across varying code graphs.

\begin{table}[ht]
\centering
\fontsize{10}{12}\selectfont
\setlength{\tabcolsep}{5pt} 
\renewcommand{\arraystretch}{1.5} 
\resizebox{\columnwidth}{!}{
    \begin{tabular}{ccccc}
    \toprule
    \makecell{\textbf{Code} \\ \textbf{Graphs}} & \makecell{\textbf{Total Security-} \\ \textbf{Sensitive Lines}} & \makecell{\textbf{Identified Security-} \\ \textbf{Sensitive Lines}} & \makecell{\textbf{Missed Security-} \\ \textbf{Sensitive Lines}} & \makecell{\textbf{Overall Identified} \\ \textbf{Percentage}} \\
    \midrule
    358 & 3088 & 2784 & 304 & 90.15\% \\ 
    \bottomrule
    \end{tabular}}
    \caption{\sys's performance on secure sensitive lines}
    \label{tab:security_critical_lines}
    \vspace{-0.5em}
\end{table}
\subsubsection{\sys Performance at Function Level}
The function-level analysis aims to evaluate \sys's ability to detect entire functions as security-sensitive based on line-level predictions. 
By automatically identifying and flagging these entire functions as secure-sensitive, we help the developers in determining which functions should be relocated to TEE. 
Therefore, the performance of \sys at the function level is highly critical. 

\Cref{tab:calgnn_performance_granularity} demonstrates \sys'ability  to identify most security-sensitive functions, achieving a higher recall of 96.23\% more than line and node level. 
However, similar to line level evaluation, we use aggregation approach to flag entire function as secure sensitive even when a single line is classified as secure sensitive by the \sys. 
This leads to more false positives resulting in lower precision than node level. 
Nevertheless, this slight aggressiveness in  flagging secure sensitive functions helps in reducing the back-and-forth switching between TEE and non-TEE environments.
\sys also achieved an F1 score of $75.23\%$ still showing its balance in minimizing both false negatives and false positives.

Furthermore, \Cref{tab:security_critical_functions} presents a summary of \sys's performance in identifying security-sensitive functions across $358$ code graphs. Out of $1237$ total functions, $1225$ were rightly identified as security-sensitive, resulting in a higher identification rate.

\sys is the first approach that produces this level of accurate and automatic identification of sensitive code for TEE integration. 

\begin{table}[ht]
\centering
\fontsize{10}{12}\selectfont
\setlength{\tabcolsep}{5pt}
\renewcommand{\arraystretch}{1.5} 
\resizebox{\columnwidth}{!}{
    \begin{tabular}{ccccc}
    \toprule
    \makecell{\textbf{Code} \\ \textbf{Graphs}} & \makecell{\textbf{Total Security-} \\ \textbf{Sensitive Functions}} & \makecell{\textbf{Identified Security-} \\ \textbf{Sensitive Functions}} & \makecell{\textbf{Missed Security-} \\ \textbf{Sensitive Functions}} & \makecell{\textbf{Overall Identified} \\ \textbf{Percentage}} \\
    \midrule
    358 & 1237 & 1133 & 104 & 91.59\% \\ 
    \bottomrule
    \end{tabular}}
    \caption{\sys's performance on secure sensitive functions}
    \label{tab:security_critical_functions}
    \vspace{-0.5em}
\end{table}

\subsection{\sys Robustness}
This section examines the robustness of \sys by evaluating its performance across different software sizes and varying project contexts. 

\subsubsection{\sys Performance Across Different Software Sizes}
\begin{figure}[ht]
    \centering
    \includegraphics[width=0.80\columnwidth]{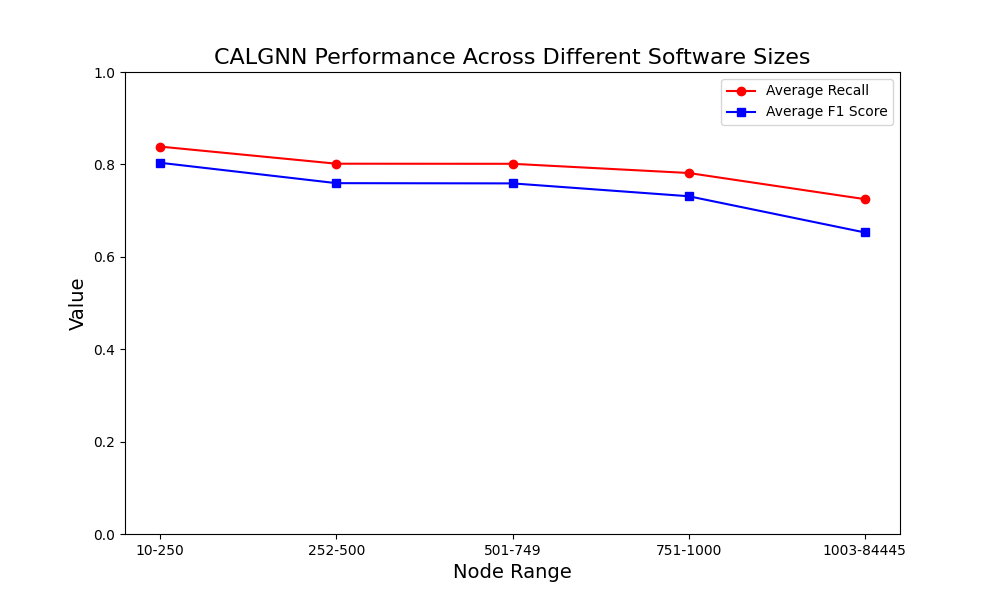} 
    \caption{\sys Performance Across Different Software Sizes}
    \label{fig:calgnn_software_sizes}
    \vspace{-0.5em}
\end{figure}

This subsection demonstrates the robustness and scalability of \sys across various software projects. 
Different sizes of source files from the constructed dataset are divided into subsets based on the number of nodes to ensure an unbiased evaluation. 
The subsets are categorized as follows: $10-250$ nodes, $252-500$ nodes, $501-749$ nodes, $751-1000$ nodes, and $1003-84445$ nodes. 
Evaluating its performance on these varying sizes of source code graphs ensures \sys's applicability in various real-world scenarios.
The performance metrics reported in~\Cref{fig:calgnn_software_sizes}, are evaluated at the node level, and several key observations can be drawn. 
Specifically, for code graphs with a smaller number of nodes, \sys achieved the highest average recall and F1 scores.

This indicates that \sys can effectively identify security-sensitive nodes in limited data contexts. 
For larger code graphs, while there is a slight decrease in recall and F1 score, the metrics still remain considerably high, showcasing the model's capability to handle substantial amounts of data without significant performance degradation.
The slight performance dip in larger code graphs can be attributed to the increased complexity and volume of subtle information present in more extensive code graphs, which poses a greater challenge for precise identification.
Nevertheless, \sys continues to demonstrate its effectiveness across varying codebase sizes, achieving nearly a size-agnostic performance that underscores its robustness and adaptability to different code structures and complexities. 

\subsubsection{\sys Robustness Across Various Software Project Contexts}

\begin{table}[ht]
\centering
\fontsize{14}{16}\selectfont 
\setlength{\tabcolsep}{10pt} 
\renewcommand{\arraystretch}{1.8} 
\resizebox{0.95\columnwidth}{!}{
    \begin{tabular}{c c c c c c}
    \toprule
    \makecell{\textbf{Software} \\ \textbf{Context}} & \textbf{Line of Code (LOC)} & \textbf{Recall (\%)} & \textbf{F1 Score (\%)} & \textbf{Precision (\%)} & \textbf{Accuracy (\%)} \\
    \midrule
    \makecell{Cryptographic Algorithms \& \\ Implementations} & 2750 & 95.15 & 78.63 & 67.33 & 92.62 \\
    \midrule
    \makecell{OpenSSL Integration \\ and Extensions} & 2875 & 95.62 & 79.14 & 77.89 & 78.53 \\
    \midrule
    \makecell{Security Tools \\ and Libraries} & 2652 & 92.17 & 79.22 & 69.25 & 88.26 \\
    \midrule
    \makecell{Networking and Secure \\ Communication} & 2254 & 97.73 & 80.58 & 72.43 & 94.72 \\
    \midrule
    \makecell{Application Development and \\ Platform-Specific Implementations} & 2567 & 88.40 & 81.86 & 78.08 & 91.15 \\
    \bottomrule
    \end{tabular}
}
\caption{\sys's Performance Across Different Software Project Contexts}\vspace{-0.5em}
\label{tab:CAL_software_context}
\end{table} 
This subsection evaluates the adaptability of \sys across diverse software project contexts. \Cref{tab:CAL_software_context} summarizes the performance metrics of \sys at the node level across different projects, demonstrating its versatility.
In contexts such as Cryptographic Algorithms and Implementations, and OpenSSL Integration and Extensions, \sys achieves high recall rates of $95.15\%$ and $95.62\%$, with corresponding F1 scores of $78.63\%$ and $79.14\%$, showing a strong ability to capture security-sensitive elements effectively.
For Networking and Secure Communication, \sys balances recall $97.73\%$ with precision $75.91\%$, leading to a similar F1 score of $80.58\%$, suggesting effective identification of critical elements in network-level security as well.
In Application Development and Platform-Specific Implementations, \sys achieves a recall $88.40\%$ with an F1 score of $81.86\%$, indicating moderate variability but still demonstrating adaptability.
Overall, \sys performs consistently well across different contexts, particularly excelling in cryptographic and security-focused areas, and shows suitability for a wide range of software contexts with minimal differences in performance.

\subsection{Case Study: Bitcoin Utility}
To showcase the capabilities of the \sys in real-world projects, we examine a Bitcoin utility tool called \textit{btcsigning}. This tool is a simplified version of the Bitcoin CLI tool provided by libbtc~\cite{libbtc}. Both btcsigning and libbtc use OpenSSL along with its extension, libsecp256k1, for ECDSA operations on the secp256k1 elliptic curve.

The \textit{btcsigning} tool has access to a transaction signing key in Wallet Import Format (WIF), which allows it to sign arbitrary messages using that key, including genuine Bitcoin transactions. The exposure of such a key could lead to substantial financial losses, emphasizing the importance of protecting it securely via TEE.

\begin{table}[ht]
    \centering
    \begin{adjustbox}{width=\columnwidth,center}
    \begin{tabular}{H>{\ttfamily\footnotesize}p{0.89\linewidth}c}
        \toprule
        Line & Code & Prob. \\ 
        \midrule
        \rowcolor{lightgray}
        262 & \lstinline|unsigned char private_key_hex[32];| & 0.9823 \\
        
        265 & \lstinline|if (decode_wif_to_hex(wif_key, private_key_hex) != 32) {| & 0.9871 \\
        
        \rowcolor{lightgray}
        266 & \quad \lstinline|printf("Invalid WIF key\n");| & 0.9784 \\ 
        
        267 & \quad \lstinline|return 1;| & 0.9851 \\
        
        \rowcolor{lightgray}
        271 & \lstinline|char un_base58_address[34];| & 0.9269 \\
        
        272 & \lstinline|bitcoin_getaddress(un_base58_address, private_key_hex);| & 0.9816 \\
        
        \bottomrule
    \end{tabular}
    \end{adjustbox}
    \caption{Converting signing key}
    \label{tab:convert_key}
    \vspace{-0.5em}
\end{table}

\Cref{tab:convert_key} shows code, where the private signing key is decoded from WIF format to raw binary format that is used by libsecp256k1, and the raw key is written into unsigned char array $private\_key\_hex$. Next, this private key is used to derive the Bitcoin address associated with the private key. The right column of the table shows the probabilities of these critical code lines. As we can see from the table, CAL correctly predicts that the code lines are security-critical in 92-98\% probability. 

\begin{table}[ht]
    \centering
    \begin{adjustbox}{width=\columnwidth,center}
    \begin{tabular}{H>{\ttfamily\footnotesize}p{0.89\linewidth}c}
        \toprule
        Line & Code & Prob. \\ 
        \midrule
        \rowcolor{lightgray}
    292 & \lstinline|unsigned char bitcoin_sig[65];| & 0.9946 \\
    293 & \lstinline|int bitcoin_sig_len = bitcoin_sign(bitcoin_sig, hash2, private_key_hex);| & 0.9874 \\
        \rowcolor{lightgray}
    294 & \lstinline|if (bitcoin_sig_len < 0) {| & 0.9854 \\
    295 & \quad \lstinline|printf("Error signing the message\n");| & 0.9938 \\
        \rowcolor{lightgray}
    296 & \quad \lstinline|return 1;| & 0.9940 \\
    297 & \lstinline|}| &  \\
        \rowcolor{lightgray}
    299 & \lstinline|printf("bitcoin_sig: ");| & 0.9955 \\
    300 & \lstinline|for (int i = 0; i < 65; i++) {| & 0.9969 \\
        \rowcolor{lightgray}
    301 & \quad \lstinline|printf("%02x", bitcoin_sig[i]);| & 0.9967 \\
    302 & \lstinline|}| & \\
        \rowcolor{lightgray}
    303 & \lstinline|printf("\n");| & 0.9962 \\
    305 & \lstinline|// Encode the signature in base64| & \\
        \rowcolor{lightgray}
    306 & \lstinline|char *signature_base64 = base64_encode(bitcoin_sig, bitcoin_sig_len);| & 0.9955 \\
    307 & \lstinline|printf("signature_base64: %s\n", signature_base64);| & 0.9941 \\
        
        \bottomrule
    \end{tabular}
    \end{adjustbox}
    \caption{Signing operation} 
    \label{tab:bitcoin_sign}
    \vspace{-0.5em}
\end{table}

Once the Bitcoin address is generated, the next crucial step involves signing the message using the $private\_key\_hex$, as shown in \Cref{tab:bitcoin_sign}. In this step, the private key is used to sign the message's hash value. \sys correctly assesses the probability as high (98.7\%). 
However, as indicated in the table, \sys mistakenly evaluates the actual signature value as security-critical, even though this value is public in the Bitcoin context. In other scenarios, such as when a verifier may accept the same signature multiple times, the signature value could be considered secret. This does not apply to Bitcoin, where the system is designed to prevent double-spending. Therefore, it is reasonable that \sys, not being trained to understand the specific logic of Bitcoin, tends to classify code as security-critical more aggressively due to the requirements of other use cases.

\begin{table}[ht]
    \centering
    \begin{adjustbox}{width=\columnwidth,center}
    \begin{tabular}{H>{\ttfamily\footnotesize}p{0.89\linewidth}c}
        \toprule
        Line & Code & Prob. \\ 
        \midrule
        \rowcolor{lightgray}
    165 & \lstinline|void double_sha256(const unsigned char *input, size_t length, unsigned char *output) {| & 0.6615 \\
    166 & \quad \lstinline|unsigned char hash[SHA256_DIGEST_LENGTH];| & 0.0788 \\
        \rowcolor{lightgray}
    167 & \quad \lstinline|SHA256(input, length, hash);| & 0.0365 \\
    168 & \quad \lstinline|SHA256(hash, SHA256_DIGEST_LENGTH, output);| & 0.5006 \\
        \rowcolor{lightgray}
    169 & \lstinline|}| &  \\
        \bottomrule
    \end{tabular}
    \end{adjustbox}
    \caption{Non-security critical code} 
    \label{tab:double_hash}
    \vspace{-0.5em}
\end{table}

\Cref{tab:double_hash} demonstrates that CAL exhibits good precision, correctly assigning lower probability scores to keyless hash operations. Although \sys assigns a lower probability to these operations, the score remains above the 0.5 threshold for hashing operations, which aligns with the security-critical definition outlined in \Cref{sec:Phase 1}.

\subsection{Runtime Performance of \sys}
The runtime performance of \sys was evaluated by measuring the time taken for various stages of processing, including code graph construction and CAL inference. 
As shown in~\Cref{tab:runtime_performance}, the total processing time for a source file of $347$ lines was,  on average, $371.36$ seconds, with a standard deviation of $2.26$ seconds.
The results are presented in~\Cref{tab:runtime_performance}. 
The feature engineering pipeline, which includes extracting complex features such as long-range dependencies on the fly, accounts for the majority of this time at $365.69$ seconds. 
Code graph construction, model loading, and inference are relatively faster, taking $6.99$, $0.214$, and $0.078$ seconds, respectively.
The longer feature engineering duration is justified by the need to provide high-quality features crucial for effective security-sensitive code identification.

\begin{table}[h!]
\centering
\renewcommand{\arraystretch}{2.3} 
\fontsize{17}{19}\selectfont 
\resizebox{\columnwidth}{!}{
\begin{tabular}{|c|c|c|c|c|c|}
\hline
\multirow{2}{*}{\textbf{Source File}} & \multicolumn{2}{c|}{\textbf{Code Graph Construction}} & \multicolumn{2}{c|}{\textbf{CAL Inference}} & \multirow{2}{*}{\textbf{Total Time (s)}} \\ \cline{2-5}
 & \textbf{Code Graph (s)} & \textbf{Feature Eng. (s)} & \textbf{Model Loading (s)} & \textbf{Inference (s)} &  \\ \hline
347 lines & 6.99 & 365.69 & 0.214 & 0.078 & 371.36 \\ \hline
\end{tabular}
}
\caption{Runtime Performance of \sys}
\label{tab:runtime_performance}
\end{table}

\section{Related Work}
\label{sec:related_work}
This section provides an overview of related literature. Notably, there is no directly relevant work addressing the specific problem that CAL aims to solve. 

We discuss the most closely related work to \sys, specifically literature that focuses on simplifying and automating the migration of applications to TEEs and GNN-based vulnerability detection methods.

\vspace{0.2cm}\noindent\textbf{Tooling for TEE Migrations:}\label{sec:related_work:tee_migrations}
Tools like Graphene~\cite{tsai2017graphene}, Haven~\cite{Haven}, and SCONE~\cite{SCONE} enable the migration of entire software environments into secure enclaves. Graphene and Haven facilitate legacy applications, while SCONE targets containerized microservices. Although these tools simplify TEE integration, they focus on full migration rather than minimizing the TCB, contrasting with \sysNoSpace's approach.

Other tools have explored optimizing TEE boundaries, similar to \sys, but require significant developer interaction. TEE-DRUP~\cite{liu2020reducing} helps developers partition applications by insourcing security-sensitive variables based on natural language heuristics, yet its reliance on manual curation makes it less efficient. SOAAP~\cite{gudka2015clean} targets multi-compartment technologies like CHERI~\cite{CHERI} and relies heavily on developer-provided annotations, making it time-consuming compared to the fully automated approach of \sys that works directly on unannotated source code.

\vspace{0.2cm}\noindent\textbf{Graph-Based Vulnerability Detection:}\label{sec:related_work:gnns}
GNNs have emerged as a leading approach in software vulnerability detection due to their capability to capture complex code structures.

Siow \etal~\cite{siow2022learning} showed that graph-based methods outperform other representations in tasks like vulnerability detection and code classification. Zhou \etal~\cite{Devign} introduced Devign, combining ASTs with control and data flow representations. Cao \etal~\cite{cao2021bgnn4vd} developed BGNN4VD, which integrates ASTs, CFGs, and data flow graphs (DFGs) into a unified Code Composite Graph (CCG). Cheng \etal~\cite{DeepWukong} presented DeepWukong, embedding semantic features along with programming logic to improve C/C++ vulnerability detection.
Recent methods have expanded upon these concepts. Hin \etal~\cite{hin2022linevd} introduced LineVD, focusing on statement-level vulnerability detection using program dependence graphs (PDGs) and a transformer-based model. Zhang \etal~\cite{zhang2023dshgt} proposed DSHGT, leveraging heterogeneous graph networks with CPGs. Li \etal~\cite{UnifiedCPG} developed UCPGVul, which combines function call graphs, CPGs, and natural code sequences. CPVD by Zhang \etal~\cite{CPVD} merges GAT with domain adaptation using CPGs, while VulChecker by Mirsky \etal~\cite{VulChecker} utilizes enriched (ePDGs). Ganz \etal~\cite{PAVUDI} introduced PAVUDI, focusing on vulnerability detection in software patches with GNNs.

While these approaches share similarities with \ourname in their use of GNNs and, in some cases, CPGs, they focus on vulnerability detection, which primarily targets small, statement-level vulnerabilities. In contrast, \ourname aims to detect security-sensitive code, requiring a broader understanding of relationships among all project components. Thus, these works are orthogonal to \ourname's objectives

In summary, CAL is pioneering in addressing the identification of security-sensitive code, an area not covered by current research, making it a foundational contribution to the fields of TEE migration and secure code analysis. None of the works in the three discussed categories are directly relevant or can be directly compared to CAL. First, while existing methods for automating application migration to TEEs share some objectives, they lack a focus on identifying security-sensitive code. Second, GNN-based vulnerability detection approaches, though similar in using GNNs for code analysis, diverge significantly in scope and methods from CAL’s specialized focus. 
\section{Conclusion}
\label{sec:conclusion}

In conclusion, \sys provides an effective solution for addressing the challenges of TEE integration by automating the identification of security-sensitive code, thereby reducing the trusted computing base and enhancing overall system security. By leveraging a graph-based approach and a custom GNN model, CAL achieves high accuracy in code isolation, minimizing the burden of manual analysis and optimizing TEE usage. Our evaluation results demonstrate CAL's potential to significantly improve the security of applications while maintaining efficiency, making it a valuable tool for developers seeking to utilize TEEs effectively.

\bibliographystyle{IEEEtran}
\bibliography{bib}
\section{Appendix}
\label{sec:appendix}
This appendix provides additional insights into the dataset, mapping strategies, and ablation studies of the development and evaluation of \sys.
\subsection{\textbf{Dataset}}

\subsubsection{Dataset Split} The dataset was initially divided into an 80\% training/validation set and a 20\% independent hold-out test set to ensure unbiased evaluation. The 80\% set was further processed using stratified $K$-Fold cross-validation (cf.~\Cref{subsec: training strategy}), maintaining consistent representation of secure-sensitive nodes across folds. 
\Cref{tab:updated_dataset} shows the size of the hold-out test set and the training/validation split, including the number of projects, code graphs, nodes, edges, and security-sensitive nodes. 
\begin{table}[h]
    \centering
    \small
    \resizebox{\columnwidth}{!}{
    \setlength\tabcolsep{3pt}
    \begin{tabular}{cccccc}
        \toprule
        \makecell{Set} & \makecell{Projects} & \makecell{Code\\Graphs} & \makecell{Nodes} & \makecell{Edges} & \makecell{Security-Sensitive\\Nodes} \\
        \midrule
        Hold-Out Test Set & 188 & 358 & 270981 & 1932224 & 53764 \\ 
        Training/Validation Set (80\%) & 516 & 1431 & 923502 & 7036398 & 323547 \\
        \bottomrule
    \end{tabular}}
    \caption{Dataset for \sys }
    \label{tab:updated_dataset}
\end{table}

\subsection{\textbf{Mapping Code Graphs to Source Code}}
Accurately mapping code graph nodes to their corresponding source code lines is crucial for identifying secure sensitive code boundaries. 
This mapping enables us to trace the predictions of the \sys model back to specific lines in the code file seamlessly. 

However, as the code graphs does not inherently contain line number information in the attributes of its nodes and edges consistently, we encounter a considerable \textit{engineering challenge} to tackle and assign the line numbers as node attributes for all the nodes to every code graph in the constructed dataset. 
For that, we developed dedicated strategies to handle cases where line number attribute is inconsistent irrespective of the node type and graph complexity, starting with: 

\noindent\textbf{Method-Based Line Number Assignment:} When a node lacks a line number as a key attribute, one of the approaches we perform is to trace the graph to find the nearest method node. 
This approach is particularly effective for nodes that are inherently linked to method elements, such as parameters, return types, or other function-related components.
By performing a BFS graph traversal~\cite{BFS}, starting from the given node, we locate and assign the line number from the nearest method node, maintaining logical consistency and alignment within the functional scope of the code.

\textbf{Proximity-Based Line Number Assignment:} For nodes lacking an inherent functional context, such as those related to various control flow, type, or structural elements, we employ a proximity-based approach to determine line numbers. Starting with a modified breadth-first search (BFS) from the target node, we examine both its successors and predecessors to identify the nearest line number within the graph. This method maintains logical closeness and consistency in representing the code structure. 

\textbf{Direct Line Number Extraction:} The third and final approach focuses on directly extracting line numbers from the code graph where they are already available as key attributes irrespective of the node type and the graph.

\subsection{Ablation Study}

In this section, we perform a detailed ablation study to evaluate the impact of various components and configurations on the performance of \sys. 
Using the best-performing \sys model as a baseline, our objective is to understand the contribution of various features and hyperparameters. 

\subsubsection{The Impact of Long Range Dependencies} 
The inclusion of long-range dependency features significantly enhances the performance of \sys, as shown in~\Cref{tab:longrange}. 
When long-range features are added, the recall improves from $81.13\%$ to $86.05\%$, indicating a better ability to identify security-sensitive nodes, which is crucial for minimizing missed detections. 
This also results in an increase in F1 score from $77.34\%$ to $81.56\%$, highlighting the importance of capturing long-range relationships within the graph for a more accurate and balanced classification.

\begin{table}[ht]
\centering
\fontsize{22}{24}\selectfont
\setlength{\tabcolsep}{11pt} 
\renewcommand{\arraystretch}{2} 
\resizebox{\columnwidth}{!}{
    \begin{tabular}{ccccccc}
    \toprule
\textbf{GNN Type} & \textbf{Features} & \textbf{Accuracy} & \textbf{Precision} & \textbf{Recall} & \textbf{F1-Score}\\
       \midrule
Gated GNN & \makecell{No Long Range} & 72.13\% & 74.12\% & 81.13\% & 77.34\% \\ 
\midrule
Gated GNN & ALL & 76.49\% & 77.50\% & 86.05\% & 81.56\% \\ 
    \bottomrule
    \end{tabular}}
\vspace{2pt}
    \caption{Impact of Long Range Features on \sys's performance}
    \label{tab:longrange}
\end{table}
\subsubsection{The Impact of Textual Features} 
The inclusion of textual features has a notable positive impact on \sys's performance, as illustrated in~\Cref{tab:textual}. 
When textual features are added, precision increases significantly from $72.29\%$ to $77.50\%$, indicating fewer false positives and more accurate identification of security-sensitive nodes. 
Additionally, the recall also slightly improves from $84.54\%$ to $86.05\%$, and the F1 score rises from $78.12\%$ to $81.56\%$, highlighting the importance of incorporating textual information.

\begin{table}[ht]
\centering
\fontsize{22}{24}\selectfont
\setlength{\tabcolsep}{11pt} 
\renewcommand{\arraystretch}{2} 
\resizebox{\columnwidth}{!}{
    \begin{tabular}{ccccccc}
    \toprule
\textbf{GNN Type} & \textbf{Features} & \textbf{Accuracy} & \textbf{Precision} & \textbf{Recall} & \textbf{F1-Score}\\
       \midrule
Gated GNN & \makecell{No Textual} & 73.78\% & 72.29\% & 84.54\% & 78.12\% \\ 
\midrule
Gated GNN & ALL & 76.49\% & 77.50\% & 86.05\% & 81.56\% \\ 
    \bottomrule
    \end{tabular}}
\vspace{2pt}
    \caption{Impact of Textual Features on \sys's performance}
    \label{tab:textual}
\end{table}

\subsubsection{The Impact of Graph Metrics as Features} 
Incorporating graph metrics such as Node Degree, Betweenness, and Closeness centralities as features provides additional indicators that help to identify secure sensitive nodes more effectively. 
As shown in~\Cref{tab:graphmetrics}, when graph metrics are added, the primary impact is on recall, which increases from $83.24\%$ to $86.05\%$. 

\begin{table}[ht]
\centering
\fontsize{22}{24}\selectfont
\setlength{\tabcolsep}{11pt} 
\renewcommand{\arraystretch}{2} 
\resizebox{\columnwidth}{!}{
    \begin{tabular}{ccccccc}
    \toprule
\textbf{GNN Type} & \textbf{Features} & \textbf{Accuracy} & \textbf{Precision} & \textbf{Recall} & \textbf{F1-Score}\\
       \midrule
Gated GNN & No Graph Metrics & 75.34\% & 76.84\% & 83.24\% & 79.91\% \\ 
\midrule
Gated GNN & ALL & 76.49\% & 77.50\% & 86.05\% & 81.56\% \\ 
    \bottomrule
    \end{tabular}}
\vspace{2pt}
    \caption{Impact of Graph Metric Features on \sys's performance}
    \label{tab:graphmetrics}
\end{table}

\subsubsection{The Impact of Edge Features}
We analyze \sys's performance with and without edge features in ~\Cref{tab:edgefeatures}.
In both settings, all node features including long-range dependencies, textual features, and graph metrics are available to isolate the specific contribution of edge features. 
The results indicate that the incorporation of edge features significantly improves the recall and F1 score, suggesting their importance in the capture of relational information within the graph data. 
Thus, the inclusion of edge features was used as a default setting in \sys's final model. 

\begin{table}[ht]
\centering
\fontsize{12}{14}\selectfont
\setlength{\tabcolsep}{8pt} 
\renewcommand{\arraystretch}{1.5} 
\resizebox{\columnwidth}{!}{
    \begin{tabular}{ccccccc}
    \toprule
    \textbf{GNN Type} & \textbf{Node Features} & \textbf{Edge Features} & \textbf{Accuracy} & \textbf{Precision} & \textbf{Recall} & \textbf{F1-Score} \\
    \midrule
    GAT & ALL & Yes & 74.79\% & 75.29\% & 86.73\% & 80.60\% \\
    \midrule
    GAT & ALL & No & 70.66\% & 74.61\% & 81.44\% & 75.01\% \\
    \bottomrule
    \end{tabular}
}
\vspace{2pt}
\caption{Impact of Edge Features on \sys’s Performance}
\label{tab:edgefeatures}
\end{table}

\subsubsection{Comparison Between Different GNN models} 
In evaluating different GNN models for \sys, we compared Gated GNN and GAT, as shown in~\Cref{tab:gnncomparison}. 
Each model has its own distinct GNN learning mechanism and was trained and evaluated under the same experimental settings. 
The Gated GNN model outperformed GATv2 in F1 score and accuracy, critical for capturing secure-sensitive patterns in complex code graphs. Its ability to manage information flow and capture long-range dependencies made Gated GNN the optimal choice for our final model.

GATv2 also performed well, especially in recall, due to its attention mechanism that adaptively focused on relevant neighbors. However, the gating mechanisms in Gated GNN proved more effective for identifying security-sensitive code.

\begin{table}[ht]
\centering
\fontsize{8}{10}\selectfont
\setlength{\tabcolsep}{3pt} 
\renewcommand{\arraystretch}{1.0} 
\resizebox{0.8\columnwidth}{!}{
    \begin{tabular}{cccccc}
    \toprule
    \textbf{GNN Type} & \textbf{Accuracy} & \textbf{Precision} & \textbf{Recall} & \textbf{F1-Score} \\
    \midrule
    GAT & 74.79\% & 75.29\% & 86.73\% & 80.60\% \\ 
    \midrule
    GatedGNN & 76.49\% & 77.50\% & 86.05\% & 81.56\% \\ 
    \bottomrule
    \end{tabular}}
\vspace{2pt}
\caption{Comparative Performance of Different GNN Models on \sys}
\label{tab:gnncomparison}
\end{table}

In summary, incorporating long-range dependencies, textual features, and graph metrics significantly improved recall and F1 scores, emphasizing their contributions in accurately identifying security-sensitive code.
Furthermore, the comparison between GNN models demonstrated the superior performance of GatedGNN, which effectively managed the information flow and captured long-range relationships, making it the optimal choice for our final model.

\end{document}